\documentclass[a4paper,twoside]{article}

\usepackage{epsfig}
\usepackage{subcaption}
\usepackage{calc}
\usepackage{algpseudocode}
\usepackage{algorithm}
\usepackage{comment}
\usepackage{amssymb}
\usepackage{amstext}
\usepackage{amsmath}
\usepackage{amsthm}
\usepackage{multicol}
\usepackage{pslatex}
\usepackage{apalike}
\usepackage{comment}
\usepackage[bottom]{footmisc}
\usepackage{longtable}
\usepackage[table,xcdraw]{xcolor}
\usepackage{threeparttable}
\usepackage{footmisc}
\usepackage{hyperref}
\usepackage{paralist}
\usepackage{wrapfig}
\usepackage{enumitem}
\usepackage{longtable}
\usepackage[table]{xcolor}
\usepackage{color}
\usepackage{amsmath}
\usepackage{multirow}
\usepackage{caption}
\usepackage{SCITEPRESS}     % Please add other packages that you may need BEFORE the SCITEPRESS.sty package.

\begin{document}

\title{Degree Centrality Algorithms For Homogeneous Multilayer Networks}

\author{\authorname{Hamza Reza Pavel\sup{1}, Abhishek Santra\sup{1} and Sharma Chakravarthy\sup{1}}
\affiliation{\sup{1}The University of Texas at Arlington}
\email{\{hamzareza.pavel,abhishek.santra\}@mavs.uta.edu, sharmac@cse.uta.edu}
}

\abstract{Centrality measures for \textit{simple graphs/networks} are well-defined and each has numerous main-memory algorithms. However, for modeling complex data sets with multiple types of entities and relationships, simple graphs are not ideal. Multilayer networks (or MLNs) have been proposed for modeling them and have been shown to be better suited in many ways. Since there are no algorithms for computing centrality measures \textit{directly} on MLNs, existing strategies reduce (aggregate or collapse) the MLN layers to simple networks using Boolean AND or OR operators. This approach negates the benefits of MLN  modeling as these computations tend to be expensive and furthermore results in loss of structure and semantics. \\
\hspace{\parindent} In this paper, we propose heuristic-based algorithms for computing centrality measures (specifically, degree centrality) on MLNs \textit{directly (i.e., without reducing them to simple graphs)} using a newly-proposed decoupling-based approach which is efficient as well as structure and semantics preserving. We propose multiple heuristics to calculate the degree centrality using the network decoupling-based approach and compare accuracy and precision with Boolean OR aggregated Homogeneous MLNs (HoMLN) for ground truth. The network decoupling approach can take advantage of parallelism and is more efficient compared to aggregation-based approaches. Extensive experimental analysis is performed on large synthetic and real-world data sets of varying characteristics to validate the accuracy and efficiency of our proposed algorithms.} %We also show that, as more information is kept in each layer, the accuracy of the decoupling-based approach increases.}


%%\onecolumn \maketitle \normalsize \setcounter{footnote}{0} \vfill

\maketitle
 
\section{Introduction}
\label{sec:introduction}
\noindent In graph-based applications, an important requirement is to measure the importance of a node/vertex, which can translate to meaningful real-world inferences on the data set. For example, cities that act as airline hubs, people on social networks who can maximize the reach of an advertisement/tweet/post, identification of mobile towers whose malfunctioning can lead to the maximum disruption, and so on. Centrality measures include degree centrality \cite{6085951}, closeness centrality \cite{2}, eigenvector centrality \cite{3}, stress centrality \cite{4}, betweenness centrality \cite{5}, harmonic centrality\cite{6}, and PageRank centrality \cite{7}, are some of the well-defined and widely-used local and global centrality measures. 

These centrality measurements use a set of criteria to determine the importance of a node or edge in a graph. Degree centrality metric measures the importance of a node in a graph in terms of its degree, which is the number of 1-hop neighbors a node has in the graph. %The degree centrality of a node is determined by its connectedness to its immediate neighbors. In many disciplines, centrality measurements are used in diverse ways. Degree centrality, for example, can be used to find the most significant person in a social network. 
Most centrality metrics are clearly defined for simple graphs or monographs, and there are numerous techniques for calculating them on simple graphs. However, for modeling complex data sets with multiple types of entities and relationships, multilayer networks have been shown to be a better alternative due to the clarity of representation, ability to preserve the structure and semantics of different types of information for the same set of nodes, and support for parallelism \cite{mln_survey,ICDMW/SantraBC17,CommFortunatoC09}.

\begin{figure}[!h]
  \centering
   {\epsfig{file = 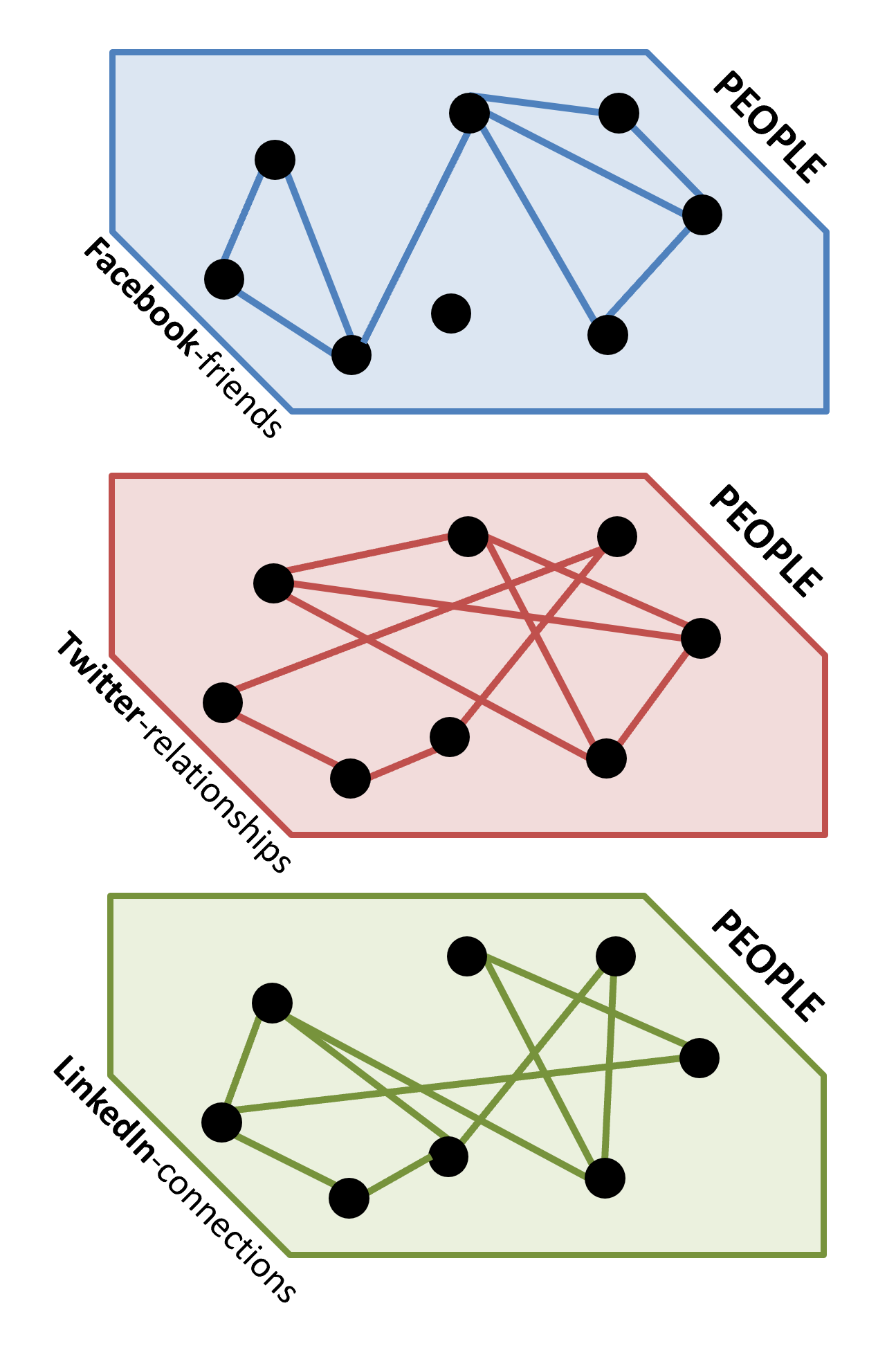, width = 5.5cm}}
  \caption{Social Media HoMLN Example}
  \label{fig:HoMLN-example}
 \end{figure}

A multilayer network~\cite{de2013mathematical} is made up of layers, each of which is a simple graph with nodes (that correspond to entities) and edges (that correspond to relationships). Nodes within a layer are connected (intra-layer edges) based on a relationship between nodes. Nodes in a layer may also be optionally connected to nodes in other layers through inter-layer edges.  As an example, the diverse interactions among the \textit{same set of people} across different social media (such as Facebook, LinkedIn, and Twitter) can be modeled using a multilayer network (see Figure~\ref{fig:HoMLN-example}.) Because the entities in each layer are the same, but the relationships in each layer are different (Facebook-friends, Twitter-relationships, LinkedIn-connections), this sort of MLN is referred to as homogeneous MLNs (or HoMLNs). As the edges between layers are implicit, they are not shown. It is also feasible to build MLNs with \textit{different types of entities and relationships} within and between layers. This form of heterogeneous MLNs (or HeMLNs) is required for modeling, for example, the DBLP data set \cite{dblpstats} with authors, articles, and conferences~\cite{mln_survey}. Hybrid Multilayer networks (HyMLNs) include both types of layers. 

For a social-network HoMLN such as the one shown in Figure~\ref{fig:HoMLN-example}, it will be interesting to find out the \textit{set of people who are the most influential in a single network or across multiple (or a subset of) social networks.} This corresponds to finding out the centrality nodes of an MLN using corresponding one or more layers. Extant algorithms that calculate centrality measures on multilayer networks are limited to simple graphs/networks and lead to the loss of structure and semantics when reduction is used. This paper presents heuristic-based algorithms for computing degree centrality nodes (or \textit{DC nodes}) on HoMLNs \textit{directly} with high accuracy/precision and efficiency. Boolean OR composition of layers is used for ground truth in this paper. 

For comparing the accuracy of the decoupling-based algorithm, we use Boolean operators for aggregation of layers and use simple graph algorithms on them for ground truth. Other types of aggregations are also possible. The aggregation of layers using AND and OR Boolean operators for homogeneous MLNs are straightforward as the nodes are the same in each layer and the operator semantics are applied to the edges. Both AND and OR operators are commutative and distributive. OR aggregation is likely to increase the size of the graph (number of edges) used for ground truth. Accuracy is computed by comparing the ground truth results for the graph with the results obtained by the decoupling-based algorithm for the layers of the same graph. The naive algorithm uses only the results of each layer and applies the operators to the results. Typically, the naive approach does not yield good accuracy requiring additional information from each layer to be retained and used for the composition algorithm using heuristics. As layers are processed \textit{independently} (may be in parallel), no information about the other layer is assumed while processing a layer.

We adapt the decoupling-based approach proposed in ~\cite{ICCS/SantraBC17,ICDMW/SantraBC17} for our algorithms. Based on this approach, we compute centrality on each layer \textit{independently once} and keep \textit{minimal} additional information from each layer for composing. With this, we can \textit{efficiently} estimate the degree centrality (DC) nodes of the HoMLN. This approach has been shown to be application independent, efficient, lends itself to parallel processing (of each layer), and is flexible for computing centrality measures on any subset of layers. The naive approach to which we compare our proposed heuristic-based accuracy retains no additional information from the layers apart from the degree centrality nodes and their values.
%%of any of the 2ˆ{n}-1 combination of AND composed layers for a $n$ layer multiplex.
Contributions of this paper are:
\begin{itemize}
\item \textbf{Algorithms} for \textit{directly} computing degree centrality nodes of Homogeneous MLNs (HoMLNs.)
\item \textbf{Several heuristics} to improve accuracy, precision, and efficiency of computed results.
\item \textbf{Decoupling-based approach} to preserve structure and semantics of MLNs.
\item \textbf{Experimental analysis} on large number of synthetic and real-world graphs with diverse characteristics.
\item \textbf{Accuracy, Precision, and Efficiency comparisons} with ground truth and naive approach.
\end{itemize}

The rest of the paper is organized as follows: Section \ref{sec:relatedwork} discusses related work. Section \ref{sec:mln_decoupling_approach} introduces the decoupling approach for MLN analysis and discusses its advantages and challenges. Section \ref{sec:degree_centrality} discusses ground truth and naive approach to degree centrality. Sections \ref{sec:DCheuristicsAccuracy}, and \ref{sec:DCheuristicsPrecision} describe composition-based degree centrality computation for HoMLNs using heuristics for accuracy and precision, respectively. Section \ref{sec:experimental_setup} describes the experimental setup and the data sets. Section \ref{sec:results} discusses result analysis followed by conclusions in Section \ref{sec:conclusions}.

\section{Related Work}
\label{sec:relatedwork}

Because complex and massive real-world data sets are becoming more popular and accessible, there is a pressing need to describe them as graphs and analyze them in various ways. However, the use of MLN for modeling poses additional challenges in terms of computing centrality measures on MLNs instead of simple graphs. Centrality measures including MLN centrality shed light on various properties of the network. Although there have been numerous studies on recognizing central entities in simple graphs, there have been few studies on detecting central entities in multilayer networks. Existing research for finding central entities in multilayer networks is \textit{use-case specific}, and there is no standard paradigm for addressing the problem of detecting central entities in a multilayer network.

Degree centrality is the most common and widely-used centrality measure. Degree centrality is used to identify essential proteins \cite{tang2013predicting}. It is also used in identifying epidemics in animals \cite{candeloro2016new} and the response of medication in children with epilepsy \cite{wang2021graph}. The most common and prominent use of degree centrality is in the domain of social network analysis. Some of the common use of degree centrality in social network analysis is identifying the most influential node \cite{7154889}, influential spreaders of information \cite{liu2016identifying}, finding opinion leaders in a social network \cite{risselada2016indicators}, etc. 

Despite being one of the most common and widely used centrality measures, very few algorithms or solutions exist to \textit{directly} calculate the degree centrality of an MLN. In this study \cite{6085951}, the author proposes a solution to find degree centrality in a 10-layer MLN consisting of the Web 2.0 social network dataset. Similar to the previous work, in~\cite{6574552}, authors identify the degree centrality of nodes using the Kretschmer method. The authors in this study \cite{yang2014predicting} proposed a node prominence profile-based method to effectively predict the degree centrality in a network. In another study \cite{gaye2016multi}, authors propose a solution to find the top-K influential person in an MLN social network using diffusion probability. More recently there has been some work in developing algorithms for MLNs using the decoupling-based approach~\cite{ICDMW/SantraBC17}.

%%SC:\abhi{5/16}{Hamza, can you highlight any other limitation of the exiting work on MLNs?}

The majority of degree centrality computation algorithms are \textbf{main memory based} and are not suitable for large graphs. They are also use-case specific. The authors adapt a decoupling-based technique proposed in \cite{ICDMW/SantraBC17} for MLNs, where each layer can be analyzed individually and in parallel, and graph characteristics for a HoMLN can be calculated utilizing the information gathered for each layer. Our algorithms follow the network decoupling methodology, which has been demonstrated to be efficient, flexible, and scalable. Achieving desired accuracy/precision/recall, however, is the challenge. Our approach is not strictly main-memory based as each layer outputs results into a file which are used for the decomposition algorithm. Also, as each layer is likely to be smaller than the OR aggregation of layers, larger size MLNs can be accommodated in our approach.

\section{Network Decoupling Approach}
\label{sec:mln_decoupling_approach}

Existing multilayer network analysis approaches convert or transform an MLN into a simple graph \footnote{A simple graph has nodes that are connected by \textit{single} edges (optionally labeled and/or directed.)}. Aggregating or projecting the network layers into a simple graph accomplishes this. Edge aggregation is used to bring homogeneous MLNs together into a simple graph. %Aggregation is performed on heterogeneous MLN using either the type-independent or projection-based aggregation methods.
Although aggregating an MLN into a simple network enables the use of currently available techniques for centrality and community discovery (of which there are many), the MLN's \textbf{structure and semantics are not retained, causing information loss}. 

\begin{figure}[!h]
  \centering
   {\epsfig{file = 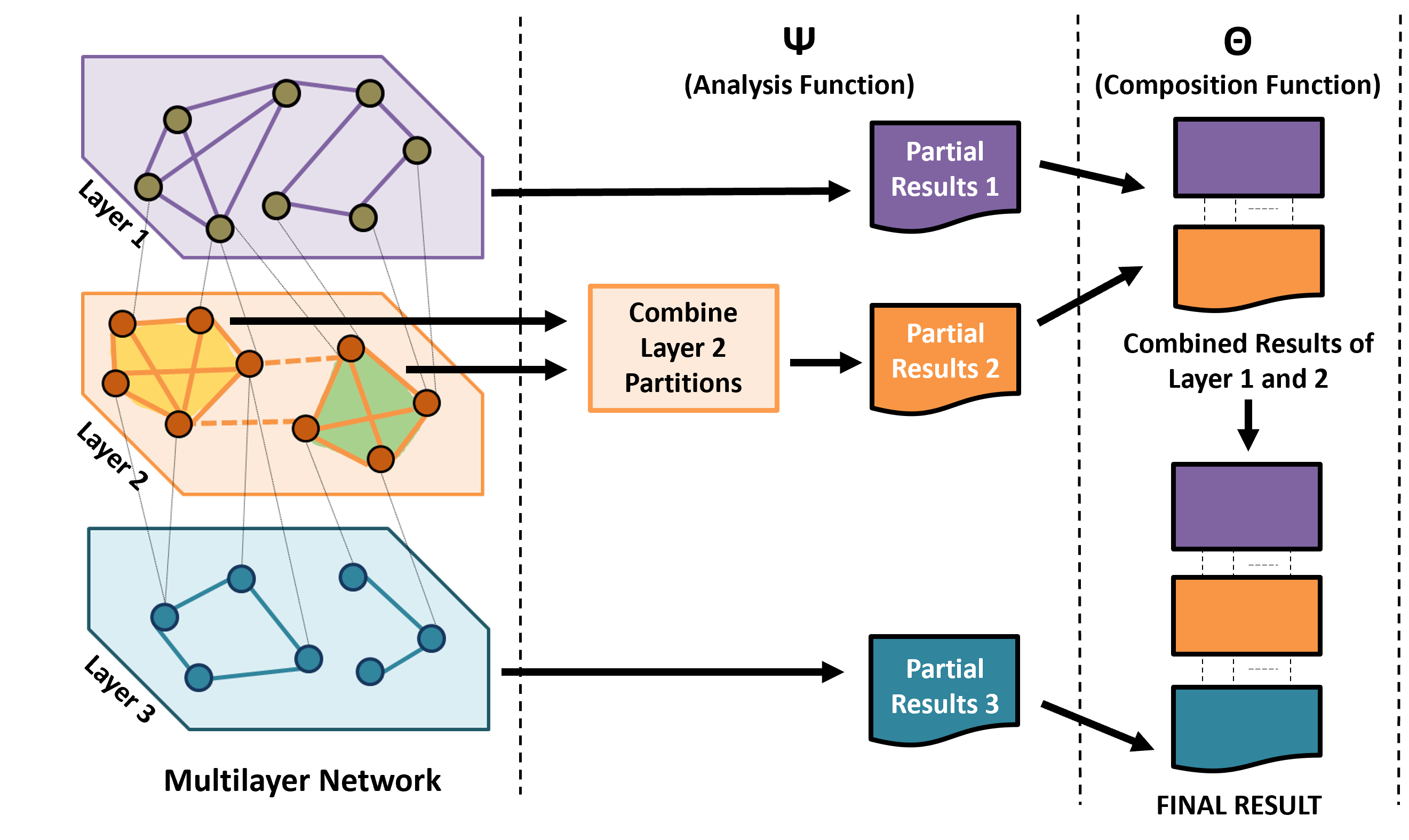, width = 7cm}}
  \caption{Overview of the network decoupling approach.}
  \label{fig:decoupling}
 \end{figure}

We use the network decoupling strategy for MLN analysis to overcome the aforementioned difficulties. Figure~\ref{fig:decoupling} shows the proposed network decoupling strategy. It entails determining two functions: one for analysis ($\Psi$) and the other for composition ($\Theta$). Each layer is analyzed independently using the analysis function (and in parallel). The partial results (as defined by the MLN) from each of the two layers are then combined using a \textit{composition function/algorithm} to obtain the HoMLN results for the two layers. MLNs with more than two layers can use this binary composition repeatedly. Independent analysis permits the use of existing techniques for each layer. Decoupling, on the other hand, increases efficiency, flexibility, and scalability along with extending the existing graph analysis algorithms to compute directly on MLNs.

As the network decoupling method preserves the structure and semantics of the data, drill-down and visualization of final results are easy to support. Each layer (or graph) is likely to be smaller, consume less memory than the whole MLN, and composition is done separately. The analysis function results are preserved and used in the composition. The requirement to recompute is reduced because the result of a single analysis function can be used by several composition functions, increasing the efficiency of the decoupling-based approach.
Individual layers can be analyzed using any of the available simple graph centrality algorithms. This method is also application-independent. As a result, the decoupling-based approach can be used to extend existing centrality algorithms to MLNs. To compose the outputs of analysis functions (partial results) into the final results, we only need to define the composition function. 

%%SC: In the earlier layer aggregation approaches, information is lost throughout the layer aggregation processes. Node and edge labels, as well as early edge connection, are examples of information loss. It is not possible to dig down on the acquired analysis result due to changes in semantics and structure in the single graph approach. We may dig down on the analysis results and acquire a deeper insight into relationships among the entities represented by the MLN using the decoupling technique because the layers are evaluated independently and the structure and semantics of the layers do not change.

%The problem with a decoupling-based approach is getting high accuracy when compared to the ground truth. Thus, 
One of the major challenges with the decoupling approach is to determine the minimum additional information to retain as part of the layer analysis step to be used during composition to improve the overall accuracy with respect to the ground truth. For many composition algorithms we have looked into, there is a trade-off between using more information from each layer and improving accuracy. This trade-off is demonstrated in this paper for one of the heuristics.

%%SC:\abhi{5/16}{Hamza, be careful with the use of terminology. there seems to be some repetition in this section. Please minimize that.}

%%SC: \abhi{5/16}{May be you remodel this section as an overview to MLNs, Boolean composition and network decoupling approach. State examples which are specific to degree centrality to make it interesting and relatable. This section should clearly give the problem statement that you are trying to address. The next section can focus only on your work on OR-based degree centrality measures for HoMLNs.}

%%SC: Both HoMLNs and HeMLNs are made up of simple graph layers. As a result, the MLN model allows for a natural splitting of a huge graph into MLN layers. In the network decoupling strategy, instead of aggregating the layers into a simple graph, we can examine the layers of an MLN individually in parallel and aggregate the results of our research.
The decoupling approach's layer-wise analysis has a number of advantages. First, only a smaller layer of the network needs to be loaded into memory, rather than the whole network. Second, the analysis of individual layers can be parallelized, reducing the algorithm's overall execution time. Finally, the composition function ($\Theta$) relies on intuition, which is built into the heuristic and takes substantially less calculation than $\Psi$.

The accuracy of an MLN analysis algorithm is determined by the information we keep (in addition to the output) during individual layer analysis. The basic minimum information we may maintain from each layer in terms of centrality measurements is the high centrality nodes of that layer, as well as their centrality values. The accuracy should potentially improve as we retain more relevant data for composition. However, determining what is relevant and should be retained to improve accuracy is the main challenge of this approach. The key hurdles are identifying the most beneficial minimal information and the intuition for their effectiveness.
\vspace{-20pt}

\section{Degree Centrality for MLNs}
\label{sec:degree_centrality}

%%\abhi{5/16}{use the term simple graph instead on single graph} dones.

%%SC:\abhi{5/16}{Also you may want to define AND, OR composition first and the need for it. This should lead to fact that algo exists for AND but not for OR.}

The degree of a node in a graph is the total number of edges that are incident on it\footnote{We use degree as 1-hop neighbors in this paper without taking direction into account. However, for directed graphs, in- or out-degree can be substituted for the heuristics proposed. Hence, we discuss only undirected graphs.}. \textit{Degree hubs are nodes in a network that have a degree larger than or equal to the network's average degree}. Degree hubs are specified for simple graphs. 
In \cite{abhishek_hubify}, the authors have proposed three algorithms to estimate degree hubs in AND composed multilayer networks. However, \textbf{there are no algorithms for calculating degree hubs for OR composed HoMLNs}.
If the HoMLN layers are composed of a Boolean operation such as OR, we can expand the notion of a hub from a simple graph to HoMLNs. In this paper, we suggest various composition functions to maximize accuracy, precision, and efficiency while estimating degree hubs in OR composed multilayer networks. 

%\abhi{5/16}{Need to properly explain what is ground truth.}

The ground truth is used to evaluate the performance and accuracy of our suggested heuristics for detecting the degree hubs of a multilayer network.

The degree centrality of a vertex $u$ in a network is defined as $C_{D}(u) = \textit{Number of adjacent neighbors}$. This value is divided by the maximum number of edges a vertex can have to normalize it. The equation for normalized degree centrality is:
\begin{equation}
\label{eq:1}
C_{D}(v) =  \frac{degree(v)}{n-1}
\end{equation}

%\abhi{5/16}{Please be consistent. Are you single greater than average OR greater than equal to average for hubs?}

High centrality hubs or degree hubs are the vertices with normalized degree centrality values higher than the other vertices. 

Even though there are different variations of degree centrality such as the group degree centrality \cite{everett1999centrality}, time scale degree centrality \cite{uddin2011time}, and complex degree centrality \cite{kretschmer2007new}, in this paper, we only address the normalized degree centrality for Boolean OR composed undirected homogeneous multilayer networks. We propose \textbf{several} algorithms to identify high centrality degree hubs in Boolean OR composed MLNs. We test the accuracy and efficiency of our algorithm against the ground truth. With extensive experiments on data sets of varying characteristics, we show that our approaches perform better than the naive approach and are efficient compared to the ground truth.

%% \abhi{5/16}{Are you proposing 4 or 5 algorithms?}
%% 4 algos. dropping dc3. so dc4 and dc5 is going to be dc3 and dc4 respectively. even though they have separate names in the data sheets.

For degree centrality, the ground truth is calculated as follows:

%%\abhi{5/16}{Use begin-itemize, end-itemize or enumerate facility given by latex, instead of hard-wiring} done

\begin{itemize}
    \item First, all the layers of the network are aggregated into a single layer using the Boolean OR composition function.
    \item Degree centrality of the aggregated graphs is calculated and the hubs are identified.
\end{itemize}

We compare the hubs computed by our algorithms against ground truth for accuracy and/or precision. We use Jaccard’s coefficient as the measure to compare the accuracy of our solutions with the ground truth.

Our aim is to design heuristics based on intuition and algorithms using the network decouple approach so that our accuracy for degree centrality is much better than the naive approach and closer to the ground truth. Efficiency is expected be better than that of the ground truth. For the naive composition approach, we estimate the degree hubs in OR composed layers as the \textbf{union} of the degree hubs of the individual layers (for OR aggregation). Even though our solution works for any arbitrary number of layers, we have focused on two layers due to its cascading effect for higher number of layers.

\section{Degree Centrality Heuristics for Accuracy}
\label{sec:DCheuristicsAccuracy}

We measure accuracy with respect to ground truth using the Jaccard coefficient. An accuracy of 1 indicates an exact match with the ground truth without any false positives or false negatives. The goal is to get accuracy as close to 1 as possible using the decoupling approach. For most applications, high accuracy is desired. 
%%SC:Apart from a few special cases, in most cases, a high accuracy from the algorithms is desired. 
In this section, we present two heuristic-based composition algorithms with better overall accuracy as compared to the naive approach.

\subsection{Degree Centrality Heuristic Accuracy 1 (DC-A1)}
%\abhi{5/16}{Need to explain, why low accuracy? Why high precision? The difference between DC2 and DC3 needs to be more clear.}

%%SC: The composition algorithms developed using heuristics DC-P1 and DC-P2 have low accuracy when compared to the naive approach, even though both algorithms have extremely high precision (further discussed in later sections).

Intuitively, with the information from each layer, we are trying to estimate the degree of a node when the layers are aggregated. If we can do it effectively, we can use the approximated average degree of the OR aggregation to determine whether a node is a hub when layers are combined. Based on the OR operator semantics, the estimated degree $estDeg_{xORy}(u)$ of a node $u$ in the OR composed layer can be $max(deg_{x}(u), deg_{y}(u))$. This happens when the one-hop neighbor of the node $u$ in layer $x$ is a subset of the one-hope neighbor of the same node in layer $y$ or vice-versa. We can use this estimated degree value of the nodes to directly calculate the degree hubs of the HoMLN in the OR composed layer. The algorithm \ref{alg:heuristic_2} describes the steps of the composition or $\Theta$ step using this heuristic. 

%%SC: Also, finding common degree hubs of the layers and finding the overlap of their neighbors (steps 2 and 3) in the algorithm \ref{alg:heuristic_1} is computationally expensive. We aim to simplify the composition algorithm to increase efficiency. As we previously discussed, 

\begin{algorithm}
\caption{Procedure for Heuristic \texttt{DC-A1}}
\label{alg:heuristic_2}
\textbf{Require:} $deg_{x}$, $deg_{y}$, $DHxORy \gets \emptyset$
\begin{algorithmic}[1]
\For{$u \in x$}
    \State $estDeg_{xORy}(u) \gets max(deg_{x}(u), deg_{y}(u))$
\EndFor
\State Calculate $DH{'}_{xORy}$ using $estDeg_{xORy}(u)$
\end{algorithmic}
\end{algorithm}

\textit{As can be seen from Table~\ref{tab:dataset1} and Figure~\ref{fig:accuracy_syn1}, this heuristic improves accuracy for data sets where the edge distribution is equal and further accuracy improves as the data set size increases. This is as expected as equal distribution of edges provides a better estimated degree for the combined layers. And for data sets with a larger number of edges, even with non-equal distribution, the average degree of the combined layers is smoother than for small data sets. This observation holds for the other synthetic data sets as well. For real-world data sets, both DC-A1 and DC-A2 are uniformly significantly better than the naive and do not deviate much from synthetic data sets with wider coverage of edge distributions and degree distributions.}

\subsection{Degree Centrality Heuristic Accuracy 2 (DC-A2)} 

In the DC-A1 heuristic, we assumed that the one-hop neighbors of a node $u$ in layer $x$ are going to be a subset of one-hop neighbors of the same node in layer $y$ or vice-versa. When we are estimating the degree of a node $u$ in the OR composed layer, there is a minimum value and maximum value for the estimated degree value of that node. The minimum of the estimated degree value is $max(deg_{x}(u), deg_{y}(u))$. Similarly, the maximum value of the estimated degree could be $min(((deg_{x}(u) + deg_{y}(u)), (n-1))$ when there is no common one-hop neighbour among layer $x$ and $y$ for node $u$. Here $n$ is the number of nodes in each layer of the HoMLN. Based on observations of various datasets, the estimated degree of a node $u$ in the OR composed layers is neither the possible minimum nor possible maximum value, rather somewhere close to the average of these values. Thus, we estimate the estimated degree of node $u$ in the OR composed layer, $estDeg_{xORy}(u)$, as the average of $max(deg_{x}(u), deg_{y}(u))$ and $min(((deg_{x}(u) + deg_{y}(u)), (n-1))$. We then use the $estDeg_{xORy}(u)$ of the nodes to calculate the degree hubs of the OR composed layer.

\textit{Note that in this heuristic, we are not using any additional information than heuristic DC-A1, but changing our estimation to a more meaningful, realistic value than taking an extreme. With this simple change in the heuristic, again from Table~\ref{tab:dataset1} and Figure~\ref{fig:accuracy_syn1}, one can see a significant improvement in accuracy over DC-A1. In fact, some of the accuracies reach as high as 0.98 which is as good as 1. One can also see that the edge distribution and data set size differences no longer have the kind of impact seen in DC-A1. also, real-world data sets match the synthetic ones to some extent.}

%%\abhi{5/16}{All proposed algorithms need to estimate the average degree of the OR-composed graph as closely as possible. Is that right? }
%% Right.
\section{Degree Centrality Heuristics for Precision}
\label{sec:DCheuristicsPrecision}

As mentioned in the previous section, we use the Jaccard coefficient as a measure of accuracy to compare the performance of our heuristics. Based on use cases, the Jaccard coefficient might not be the only measure for many real-world applications. For example, An airline is trying to expand its operation to a new city based on the air route and operation of other competitors. This problem can be modeled as a problem to find the degree hubs of a HoMLN where each node of the HoMLN is a city and each layer represent the route of the competitors among these cities. In this scenario, a high precision algorithm is preferred as a false positive in identifying a hub might lead the airline to expand to a city without much traffic and incur loss due to the expansion. Advertising on multiple social networks also has a similar need to avoid false positives. Hence, in general, it is meaningful to identify heuristics that do not produce any false positives or any false negatives depending upon the application's need. In this section, we provide two heuristic-based composition algorithms to find the degree hubs of a HoMLN with high precision. 

\subsection{Degree Centrality Heuristic Precision 1 (DC-P1)}
For the Boolean OR operator composed ground-truths, if a node is a degree hub (DH) in layer $x$ \textit{or} layer $y$, then it is likely that the node is going to be a degree hub in the OR composed ground truth. We use this intuition as the basis for heuristic DC-P1 which is used to develop the first composition algorithm for the $\Theta$ function to compute high precision degree hubs. %This algorithm is similar to algorithm 1 of \cite{abhishek_hubify} which was used to find the degree centrality hubs of AND composed ground truths. 

%%SC: \abhi{5/16}{Please be extremenly careful with the heuristics - what is the input (set or value), output, etc.? Look the revsion below and compare against the commented para. Revise the other heuristics on similar lines. Also, Algorithm 1 is for Theta, where already know NBD-x and NBD-y from the analysis phase. Why are you writing line 2 to 7. Separate the analysis algorithm and the composition algorithm.}

% EARLIER: HAMZA version
%As we previously mentioned, in the analysis function $(\Psi)$ of the decoupling approach we analyze the layers of the HoMLN and use the partial results and additional information to obtain the final results for the MLN. For DC1, after analysis $(Psi)$ phase of each layer (say layer $x$), we keep the degree hubs of that layer $DH_{x}$, the average degree of the nodes in that layer $avgDeg_{x}$, and the set of one-hop neighbors of the degree hubs $NBD_{x}$ 

% REVISED VERSION
As we previously mentioned, in the analysis function $(\Psi)$ of the decoupling approach we analyze the layers of the HoMLN and use the partial results and additional information to obtain the final results for the MLN. In DC-P1, after the analysis $(\Psi)$ phase of each layer (say layer $x$), we keep the set of degree hubs $DH_{x}$, the average degree $avgDeg_{x}$, and the set of one-hop neighbors of each degree hub (say, $u$) $NBD_{x} (u)$\footnote{This is the additional information we retain from each layer to improve precision as we have indicated earlier.}.

%%SC: \abhi{5/16}{TIP: Again avoid repetition, you will explain in detail the meaning of $\Psi$ and $\Theta$ step once in the previous section. Here just mention, why and what is the $\Psi$ and $\Theta$ step? Be specific and to the point.}
\begin{table*}[htb]
%%\vspace{-20pt}
\caption{Summary of Synthetic Data Set-1}
\label{tab:dataset1}
\scriptsize
\centering
\renewcommand{\arraystretch}{1.4}
\begin{tabular}{|c|c|c|c|c|c|}
\hline
{Base Graph} & \multirow{2}{*}{G$_{ID}$} & {Edge Dist. \%} &  \multicolumn{3}{c|}{\#Edges} \\ 
\cline{4-6}
 \#Nodes, \#Edges & & \textit{in Layers} & L1 & L2 & L1 OR L2 \\ 
\hline
\multirow{3}{*}{100KV, 500KE} & 1 & 70,30             & 350000   & 150000   & 499587           \\ 
\cline{2-6}
& 2 & 60,40 & 200000 & 300000   & 499505           \\ 
\cline{2-6}
& 3 & 50,50 & 250000 & 250000   & 499505           \\ 
\hline
\hline
\multirow{3}{*}{100KV, 1ME} & 4 & 70,30 & 700000   & 300000   & 998303           \\ 
\cline{2-6}
& 5     & 60,40 & 600000   & 400000   & 998176           \\ 
\cline{2-6}
& 6   & 50,50  & 500000   & 500000   & 997998           \\ 
\hline
\hline
\multirow{3}{*}{100KV, 2ME} & 7 & 70,30  & 600000   & 1400000   & 1993608           \\ 
\cline{2-6}
& 8 & 60,40   & 1200000   & 800000   & 1992855           \\ 
\cline{2-6}
& 9 & 50,50   & 1000000   & 1000000   & 1992207           \\ 
\hline
\hline
\multirow{3}{*}{300KV, 1.5ME} & 10    &  70,30 & 1050000   & 450000   & 1499463          \\ 
\cline{2-6}
& 11     & 60,40               & 900000   & 600000   & 1499425          \\ 
\cline{2-6}
& 12    & 50,50           & 750000   & 750000   & 1499347          \\ 
\hline
\hline
\multirow{3}{*}{300KV, 3ME} & 13       & 70,30  & 2100000   & 900000   & 2997825          \\ 
\cline{2-6}
& 14 & 60,40       & 1800000   & 1200000   & 2997627          \\ 
\cline{2-6}
& 15 & 50,50   & 1500000   & 1500000   & 2997538          \\ 
\hline
\hline
\multirow{3}{*}{300KV, 6ME} & 16  & 70,30    & 4200000   & 1800000  & 5991761          \\ 
\cline{2-6}
& 17 & 60,40   & 3600000  & 2400000   & 5990599          \\ 
\cline{2-6}
& 18 & 50,50   & 3000000  & 3000000  & 5990044          \\ 
\hline
\hline
\multirow{3}{*}{500KV, 2.5ME} & 19    & 70,30   & 1750000   & 750000   & 2499344          \\ 
\cline{2-6}
& 20 & 60,40   & 1500000   & 1000000   & 2499238          \\ 
\cline{2-6}
& 21 & 50,50   & 1250000   & 1250000   & 2499166          \\ 
\hline
\hline
\multirow{3}{*}{500KV, 5ME} & 22    & 70,30  & 3500000  & 1500000   & 4997388          \\ 
\cline{2-6}
& 23 & 60,40       & 3000000   & 2000000  & 4996910          \\ 
\cline{2-6}
& 24 & 50,50   & 2500000   & 2500000   & 4997209          \\ 
\hline
\hline

\multirow{3}{*}{500KV, 10ME} & 25   & 70,30    & 7000000  & 3000000  & 9989402          \\ 
\cline{2-6}
& 26 & 60,40   & 6000000  & 4000000  & 9989190          \\ 
\cline{2-6}
& 27 & 50,50   & 5000000  & 5000000  & 9987447          \\ 
\hline
\end{tabular}
%%\vspace{-20pt}
\end{table*}

During the $\Theta$ step, we use the stored partial results and additional information to estimate the hubs for two layers (say layer $x$ and layer $y$). As for the OR composed ground-truth graph, the number of edges for a node is likely to increase. We can estimate the average degree of the OR composed layer,  $avgEstDeg_{xORy}$, to be the maximum between $avgDeg_{x}$ and $avgDeg_{y}$. For each node present in either $DH_{x}$ or $DH_{y}$, if the union of their one-hop neighbors set is more than $avgEstDeg_{xORy}$, we consider that node a degree hub in the OR composed layer of $x$ and $y$. Algorithm \ref{alg:heuristic_1} shows the detailed steps of the composition algorithm ($\Theta$.)

\begin{algorithm}
\caption{Procedure for $\Theta$ using Heuristic DC-P1}
\label{alg:heuristic_1}
\textbf{Require:} $DH_{x}$, $avgDeg_{x}$, $DH_{y}$, $avgDeg_{y}$, $NBD_{x}$, $NBD_{y}$, $DH_{xORy} \gets \emptyset $
\begin{algorithmic}[1]
\State $avgEstDeg_{xORy} \gets max(avgDeg_{x}, avgDeg_{y})$
%\For{$u \in DH_{x}$}
%    \State $NBD_{x}(u) \gets$ one hop neighbors of u in layer x 
%\EndFor
%\For{$u \in DH_{y}$}
%    \State $NBD_{y}(u) \gets$ one hop neighbors of u in layer y 
%\EndFor
\For{$u \in DH_{x} \cup DH_{y}$}
    \If{$|NBD_{x}(u) \cup NBD_{y}(u)| >= avgEstDeg_{xORy}$}
        \State $DH{'}_{xORy} \gets DH{'}_{xORy} \cup u$
    \EndIf{}
\EndFor
\end{algorithmic}
\end{algorithm}

%\sharma{5/6/22}{plz add some intuitive explanation such as these AFTER the discussion of each heuristic. Also, indicate how much more information is retained and improvement you are getting from expt section. are the colors between naive and DC1 interchanged in Figure 3? plz check. i am assuming that for the explanation below. if that is not the case, why do we even include DC1?}
%%to-do if time and page is remaining

\textit{Degree hubs and their values for each layer allow us to compute the higher bound of the average for the aggregated graph. One-hop neighbor information is used to reduce or eliminate false positives. However, as these are retained \textit{only} for hubs, information is still not complete. Even with this limited additional information, as we will see in the experimental section (Section~\ref{sec:results}), there is a significant improvement in precision over the naive for all data sets. For the synthetic data sets, we get a precision of 100\% and for the real-world data sets we get a mean precision of 96\% (Refer to Section \ref{sec:results}).}

%\abhi{5/16}{If you are having a lot of notations in the paper, you may consider adding a table with all in the notations at one place, like deg, NBD, etc.}

\subsection{Degree Centrality Heuristic Precision 2 (DC-P2)}

%%\abhi{5/16}{Please read this carefully. Not clear in the current format. Also, what is the intution behind the estimated average degree, why max?}
%%for OR composed aggregated layers, the aggregated graph will have at least max number of edges for a node (if all the edge overlaps) or more than that. 

Based on how the edges are distributed in the layers of an MLN, the actual average degree of the OR composed ground truth, $avgDeg_{xORy}$, of layers $x$ and $y$ might differ from the estimated $avgEstDeg_{xORy}$ in DC-P1. If the $avgEstDeg_{xORy}$ is substantially greater than $avgEstDeg_{xORy}$, then a lot of nodes will not be included as a hub in the OR composed layer despite having enough common neighbors across both layers $x$ and $y$. Similarly, if $avgEstDeg_{xORy}$ is smaller than $avgEstDeg_{xORy}$, a lot of false positives will be generated as hubs in the OR composed layer. 

To better estimate the $avgEstDeg_{xORy}$, we keep the degree of each node from each layer as additional information during the $\Psi$ step. This allows us to estimate the individual degree of a node $u$ in the OR composed layer from its degree information in layer $x$ and layer $y$. If the degree of a node $u$ in layer $x$ is $deg_{x}(u)$ and degree of the same node in layer $y$ is $deg_{y}(u)$, then estimated degree of node $u$ in the OR composed layer, $estDeg_{xORy}(u)$, is going to be $max(deg_{x}(u), deg_{y}(u))$. Using the estimated degree $estDeg_{xORy}(u)$ of each node $u$, we calculate the $avgEstDeg_{xORy}$. The rest of the steps of the algorithm are same as \ref{alg:heuristic_1}.
\vspace{-20pt}

%%\sharma{5/6/22}{need CORRECT explanation. if dark blue is DC1, this is better than that. Otherwise, it is many a times not even as good as naive. we need to provide proper explanation once the confusion is sorted out. It is not just the question of sumping results and asking the reviewer to make sense of it!}
%% as discussed in the meeting, the DC-P1 and DC-p2 provides significantly better precision compared to naive and the other two heuristics.

\section{Experimental Analysis}

\subsection{Data Sets and Environment}

\label{sec:experimental_setup}
\begin{table*}[ht]
\vspace{-20pt}
\caption{Summary of Real World Data Set}
\label{tab:dataset_realworld}
\scriptsize
\centering
\renewcommand{\arraystretch}{1.4}
\begin{tabular}{|c|c|c|c|c|c|}
\hline
{Base Graph} & \multirow{2}{*}{G$_{ID}$} & {Edge Dist. \%} &  \multicolumn{3}{c|}{\#Edges} \\ 
\cline{4-6}
 \#Nodes, \#Edges & & \textit{in Layers} & L1 & L2 & L1 OR L2 \\ 
\hline
\multirow{3}{*}{735KV, 2.6ME} & amazon-2008\_1 & 50,50             & 1306357   & 1304863   & 1958865           \\ 
\cline{2-6}
& amazon-2008\_2 & 70,30 & 1828100 & 784552   & 2063141           \\ 
\cline{2-6}
& amazon-2008\_3 & 90,10 & 2349969 & 261133   & 2376256           \\ 
\hline
\hline
\multirow{3}{*}{325KV, 1.7ME} & cnr-2000\_1 & 50,50 & 876444   & 876383   & 1314919           \\ 
\cline{2-6}
& cnr-2000\_2     & 70,30 & 1226781   & 525244   & 1384962           \\ 
\cline{2-6}
& cnr-2000\_3   & 90,10  & 1577646   & 175236   & 1595367           \\ 
\hline
\hline
\multirow{3}{*}{100KV, 1.5ME} & uk-2007-05\_1 & 50,50  & 759899   & 761252   & 1141215           \\ 
\cline{2-6}
& uk-2007-05\_2 & 70,30   & 1065435   & 455957   & 1202326           \\ 
\cline{2-6}
& uk-2007-05\_3 & 90,10   & 1369767   & 152167   & 1385013           \\ 
\hline
\hline
\end{tabular}
\vspace{-20pt}
\end{table*}
The NetworkX \cite{hagberg2008exploring} package is used in our Python implementation. The experiments were carried out on a single node SDSC Expanse \cite{xsede}. Each node in the cluster runs the CentOS Linux operating system using an AMD EPYC 7742 CPU with 128 cores and 256GB of RAM hardware.
Both synthetic and real-world data sets were used to evaluate the proposed methodologies. PaRMAT \cite{parmat}, a parallel version of the popular graph generator RMAT \cite{rmat}, which uses the Recursive-Matrix-based graph generation technique, was used to create the synthetic data sets.  
\begin{figure}[htb]
  \centering
   {\epsfig{file = 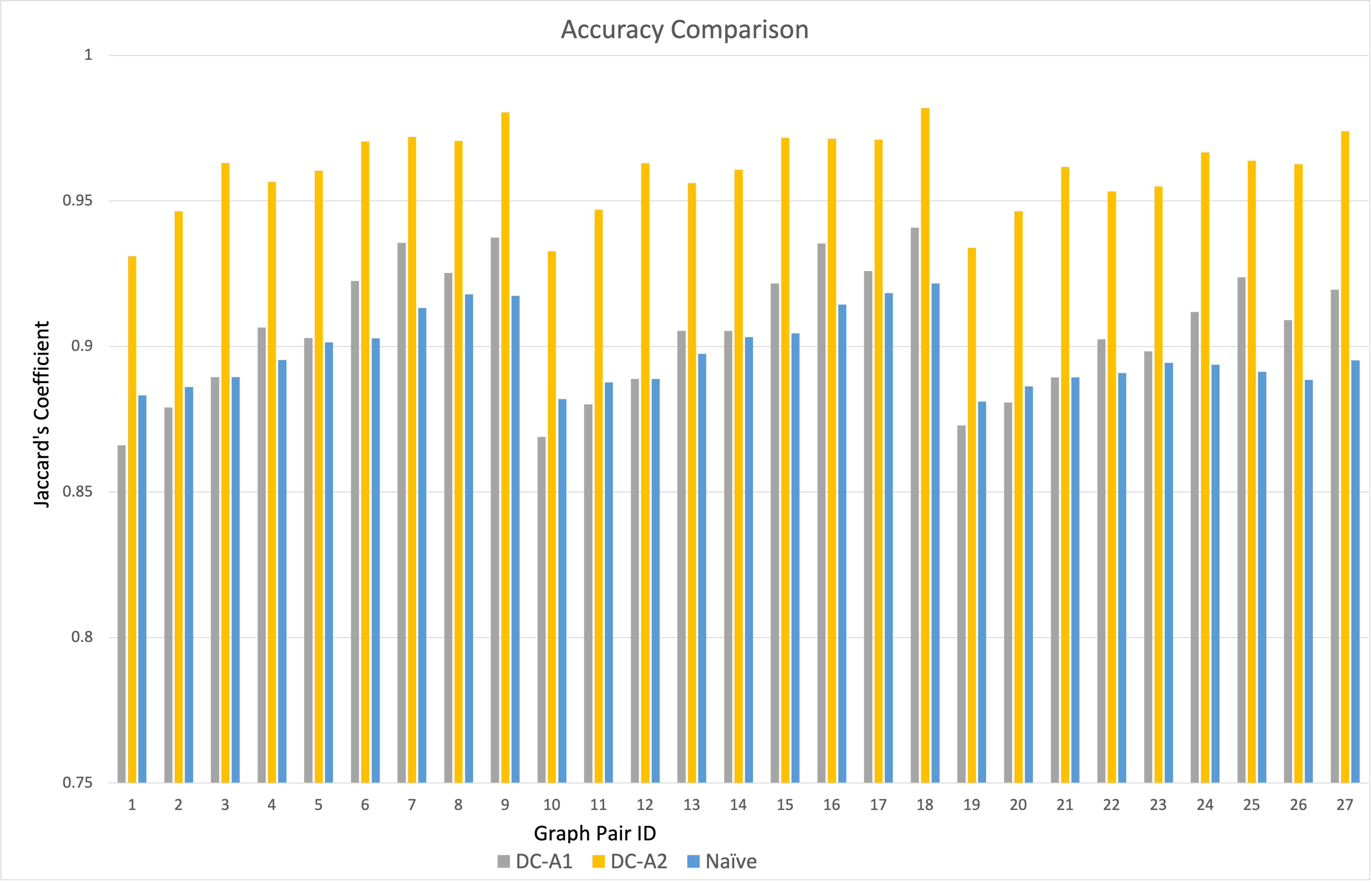, width = 0.5\textwidth}}
  \caption{Accuracy Comparison for Synthetic Dataset 1(Refer Table \ref{tab:dataset1}).}
  \label{fig:accuracy_syn1}
 \end{figure}
We use PaRMAT to produce three sets of synthetic data sets for each base graph for experimentation. Our synthetic data set consists of 27 HoMLNs with two layers, each with a different edge distribution. The base graphs start with 100K vertices with 500K edges and go up to 500K vertices and 10 million edges. In the first synthetic data set, both HoMLN layers have \textit{power-law degree distribution}. In the second synthetic data set, one layer (L1) follows \textit{power-law degree distribution} and the other one (L2) follows \textit{normal degree distribution}. In the final synthetic data set, both layers have \textit{normal degree distribution}. 
 \begin{figure}[h]
  \centering
   {\epsfig{file = 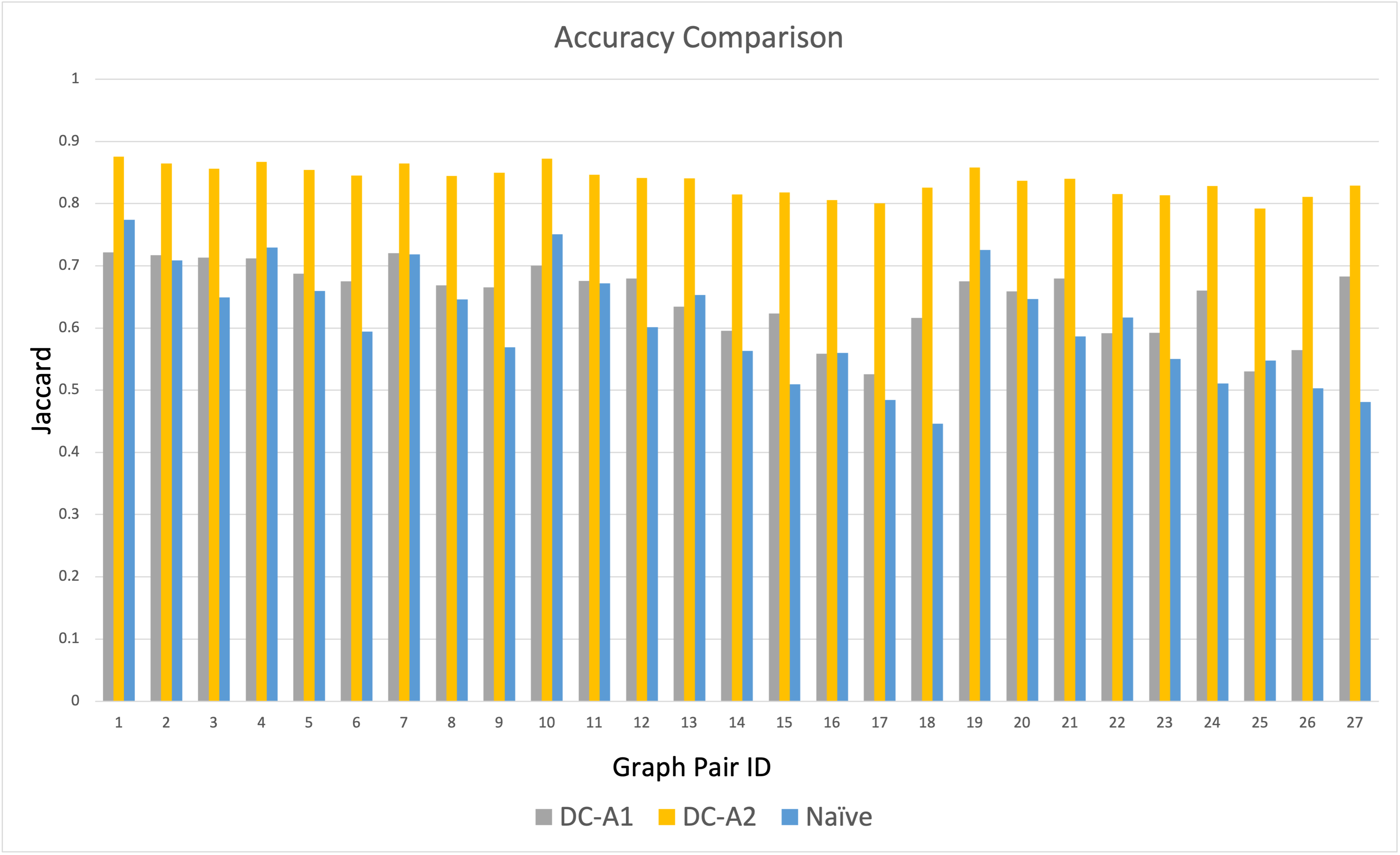, width = 0.5\textwidth}}
  \caption{Accuracy Comparison for Synthetic Dataset 2.}
  \label{fig:accuracy_syn2}
 \end{figure}
For each of the aforementioned data sets, three edge distributions (70, 30; 60, 40; and 50, 50) for a total of 81 HoMLNs with varied edge distributions, number of nodes, and edges are used for experimentation and validation of the proposed heuristics. Table \ref{tab:dataset1} shows the different 2-layer HoMLN used in our data set used in our experiments which are part of the synthetic data set 1. The synthetic data set 1 consists of HoMLN where both the layers have the power-law distribution of edges (L1: Power-law, L2: Power-law). The other two data sets, synthetic data set 2 and 3 have a similar number of nodes and edges in each layer but have (L1: Power-law, L2: Normal) and (L1: Normal, L2: Normal) edge distribution.

\begin{table*}[htb]
\vspace{-20pt}
\caption{Accuracy Improvement of \texttt{DC-A1} and \texttt{DC-A2} over Naive}
\label{tab:accuracy_summary_syn}
\scriptsize
\centering
\renewcommand{\arraystretch}{1.4}
\begin{tabular}{|c|c|c|c|c|c|}
\hline
\multirow{2}{*}{Data Set} & Degree Distribution & \multicolumn{4}{c|}{Mean Accuracy}  \\
\cline{3-6}
 & L1, L2 & \texttt{DC-A1} & \texttt{DC-A2} & \texttt{DC-A1} vs. Naive & \texttt{DC-A2} vs. Naive \\
\hline
Synthetic-1 & Power law, Power law    & 90.53\% & 96.01\% & +0.86\% & +6.96\% \\
\hline
Synthetic-2 & Power law, Normal & 64.90\% & 83.74\% & +6.48\%  & {\bf \textcolor{blue}{+37.38\%}}  \\
\hline
Synthetic-3 & Normal, Normal       & 76.14\% & 88.72\% & -4.32\% & {\bf \textcolor{blue}{+11.47\%}} \\
\hline
Real world & Power law, Power law       & 93.11\% & 98.9\% & +10.03\% & +10.92\% \\
\hline
\end{tabular}%

\end{table*}

For our real-world-like data set, the network layers are generated from real-world like monographs using a random number generator. The real-world-like graphs are generated using RMAT with parameters to mimic real-world graph data sets as discussed in~\cite{chakrabarti2005tools}. As a result, the graphs are not single connected components and neither are their ground truth graph.
  \begin{figure}[htb]
  \centering
   {\epsfig{file = 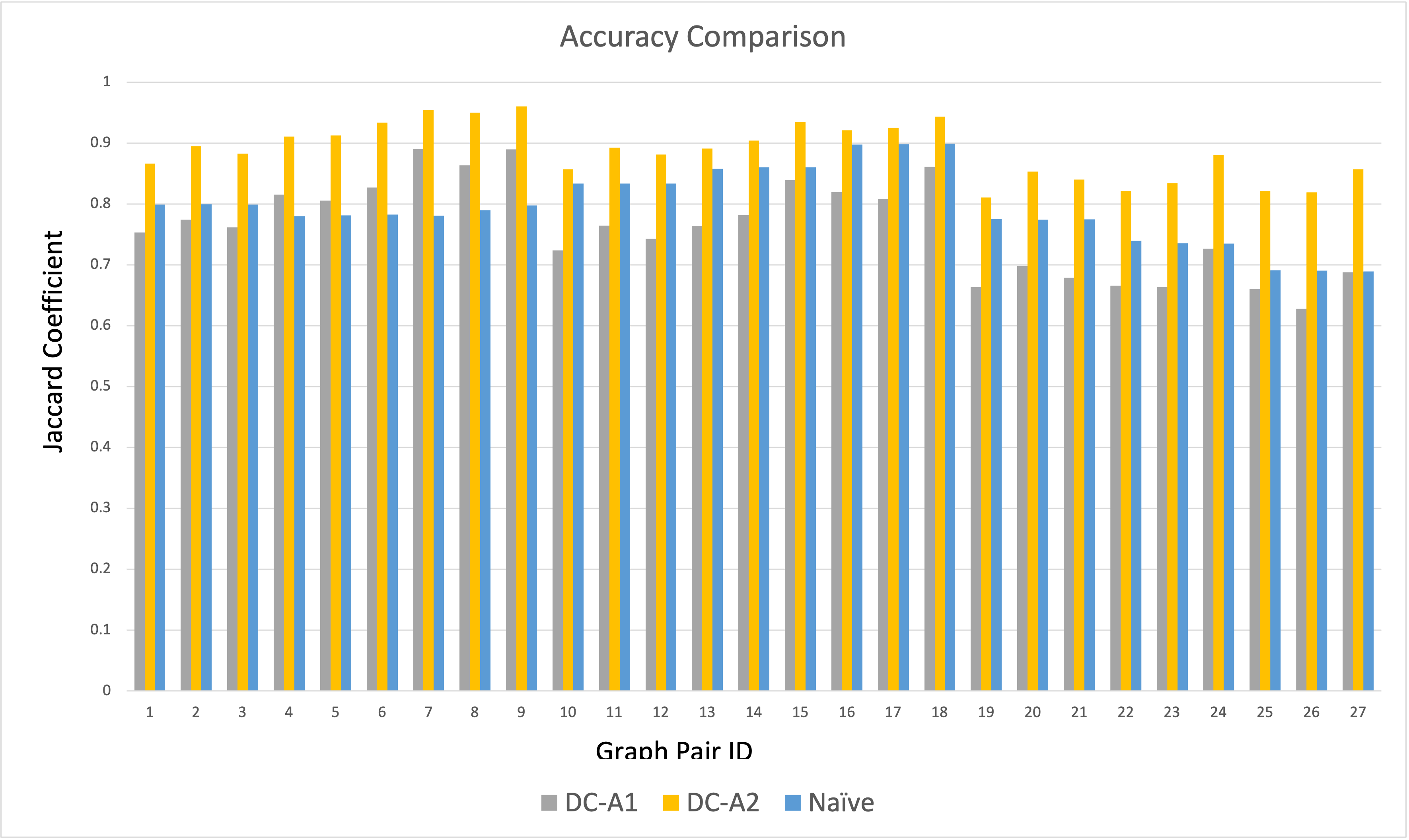, width = 0.5\textwidth}}
  \caption{Accuracy Comparison for Synthetic Dataset 3.}
  \label{fig:accuracy_syn3}
 \end{figure}

\subsection{Result Analysis and Discussion}
\label{sec:results}

\begin{figure}[h]
  \centering
   {\epsfig{file = 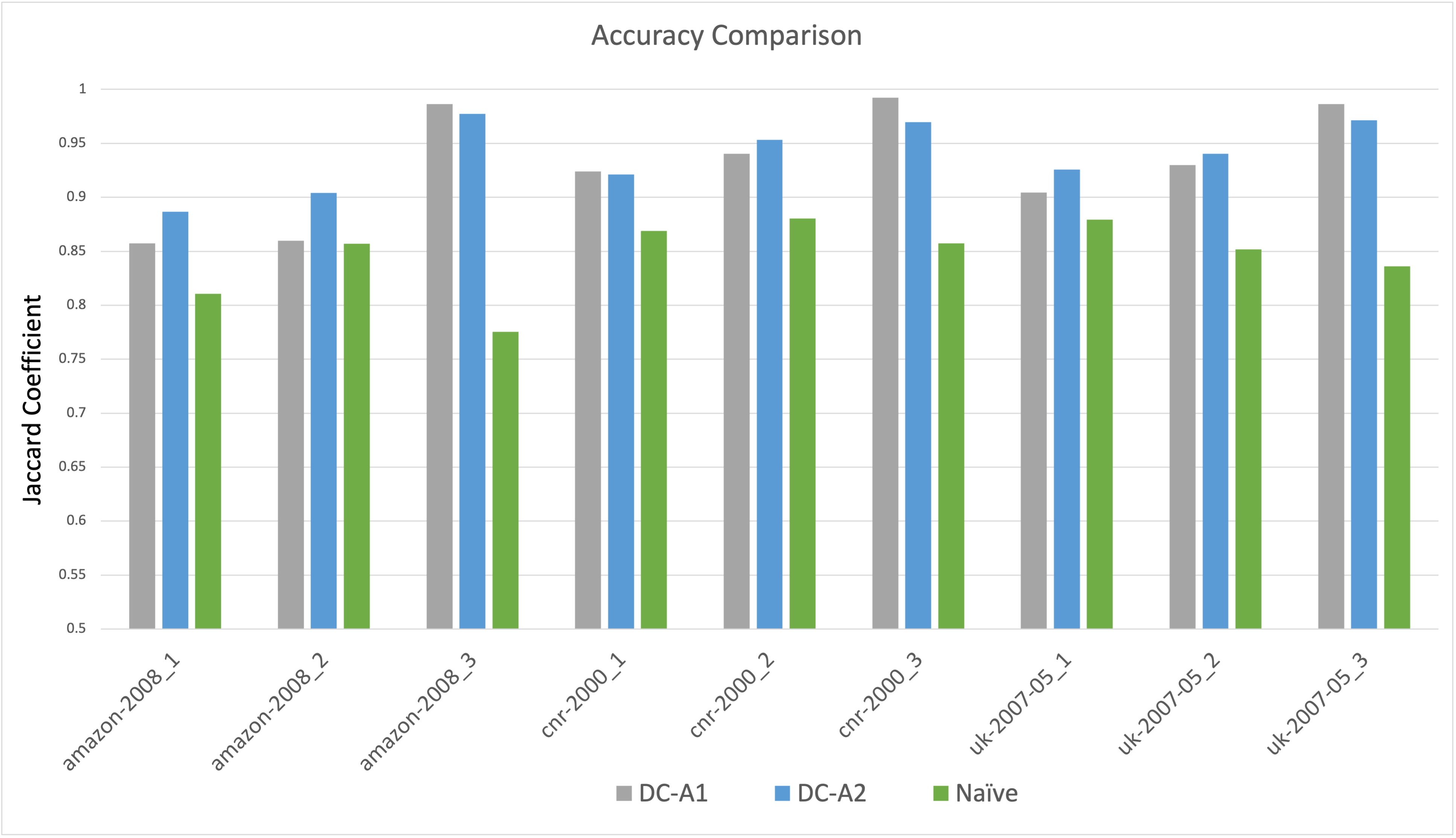, width = 0.5\textwidth}}
  \caption{Accuracy Comparison for Real World Dataset(Refer Table \ref{tab:dataset_realworld}).}
  \label{fig:accuracy_realworld}
 \end{figure}

In this section, we present our experimental results. We have tested our proposed heuristics on large real-world and synthetic datasets. As a measure of accuracy, we use the Jaccard coefficient and precision. We compare the execution time of our heuristics against the ground truth execution time as a measure of performance. 
The Figures \ref{fig:accuracy_syn1}, \ref{fig:accuracy_syn2}, and \ref{fig:accuracy_syn3} shows the Jaccard coefficient for accuracy of the proposed heuristics-based approaches DC-A1, DC-A2, and the naive approach for the synthetic dataset 1 dataset 2, and dataset 3 respectively. While calculating the Jaccard coefficient, we consider the nodes with equal to or higher than the average degree value in the ground truth as degree hubs. The heuristic DC-A2 performs the best when the accuracy metric is the Jaccard coefficient. It always shows higher accuracy than the naive approach. The heuristic DC-A1 performs better than the naive approach in most cases.
Figure \ref{fig:accuracy_realworld} shows the Jaccard's coefficient for the proposed heuristics in real-world dataset \cite{BoVWFI} \cite{BRSLLP}. Both heuristics DC-A1 and DC-A2 perform better than the naive approach for all the HoMLN in the dataset.
Table \ref{tab:accuracy_summary_syn} shows the mean accuracy and average percentage gain in accuracy for the synthetic and real-world data sets. For all data sets, DC-A2 \textit{outperforms} the naive approach. 
 The DC-A1 heuristic performs poorly when both layers have a normal distribution of edges, but performs better than naive in other cases. One reason for the low percentage gain compared to the naive approach is, that for Boolean OR aggregated HoMLN,  the naive approach itself has relatively high accuracy.
 %%SC: (compared to Boolean AND aggregated HoMLN). 
 
\begin{figure}[htb]
  \centering
   {\epsfig{file = 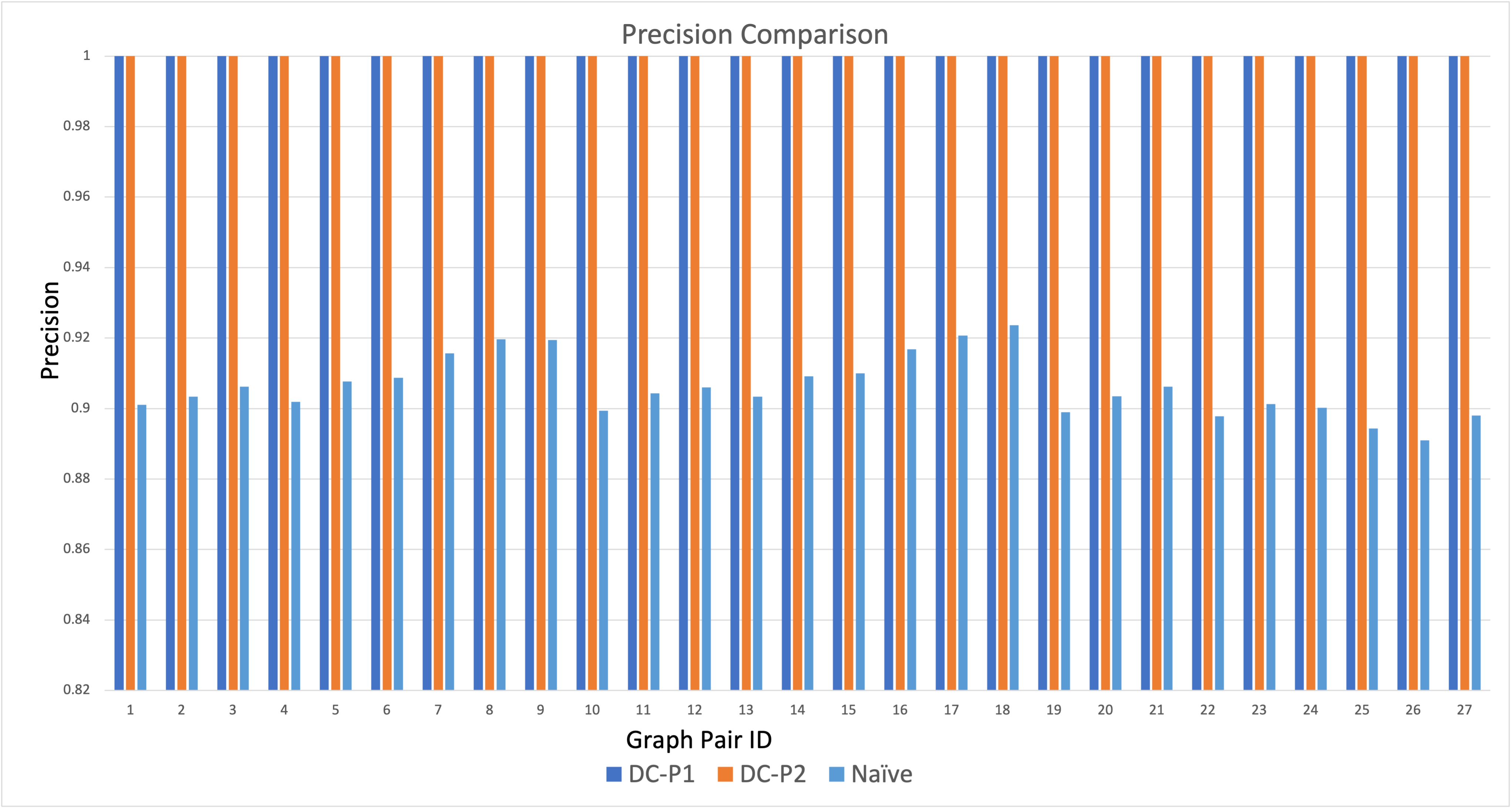, width = 0.5\textwidth}}
  \caption{Precision Comparison for Synthetic Dataset 1 (Refer Table \ref{tab:dataset1}).}
  \label{fig:precision_syn1}
 \end{figure}

\begin{figure}[htb]
\centering
{\epsfig{file = 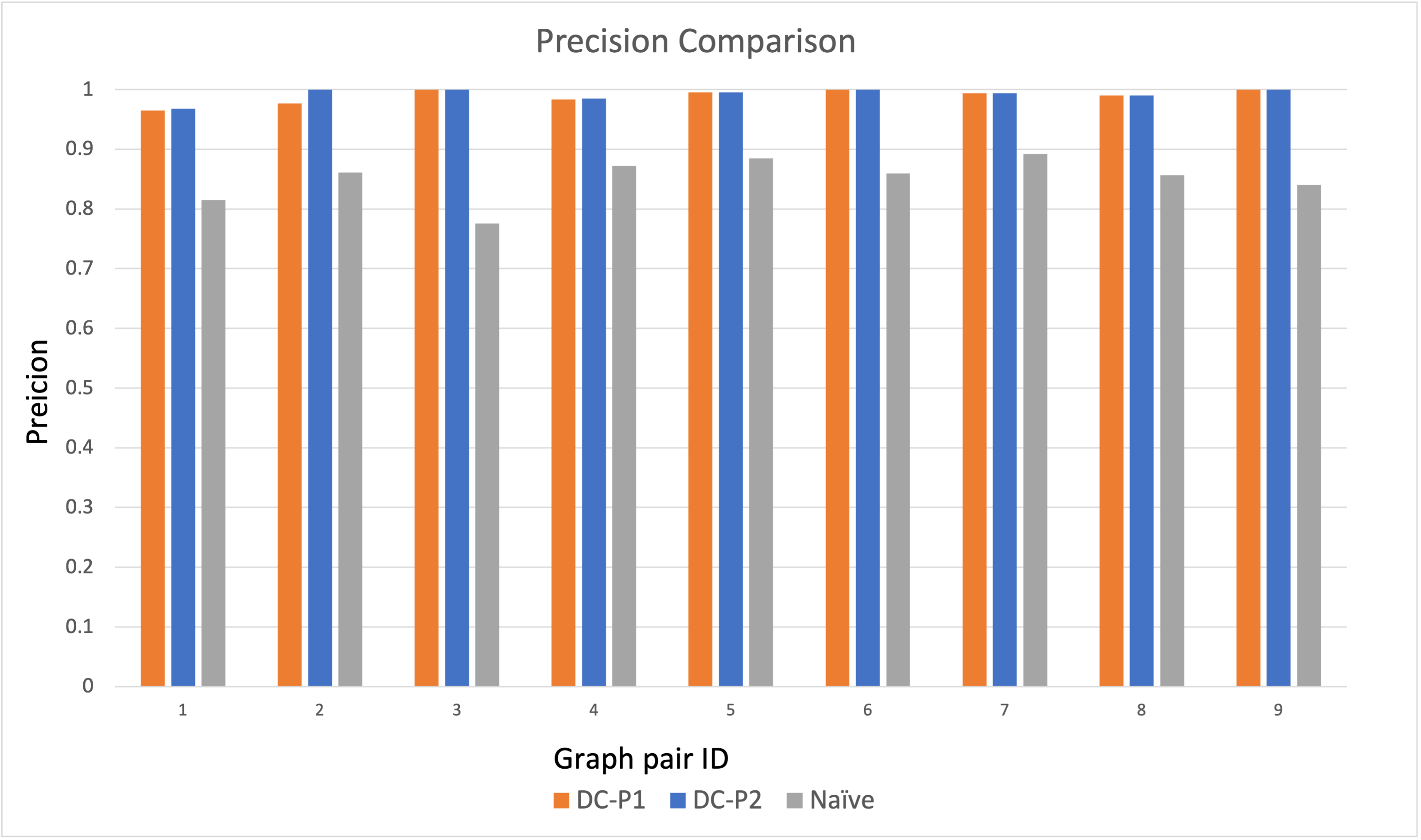, width =0.5\textwidth}}
\caption{Precision Comparison for real world dataset(Refer Table \ref{tab:dataset_realworld}).}
\label{fig:precision_realworld}
\end{figure}

\begin{figure}[htb]
\centering
{\epsfig{file = 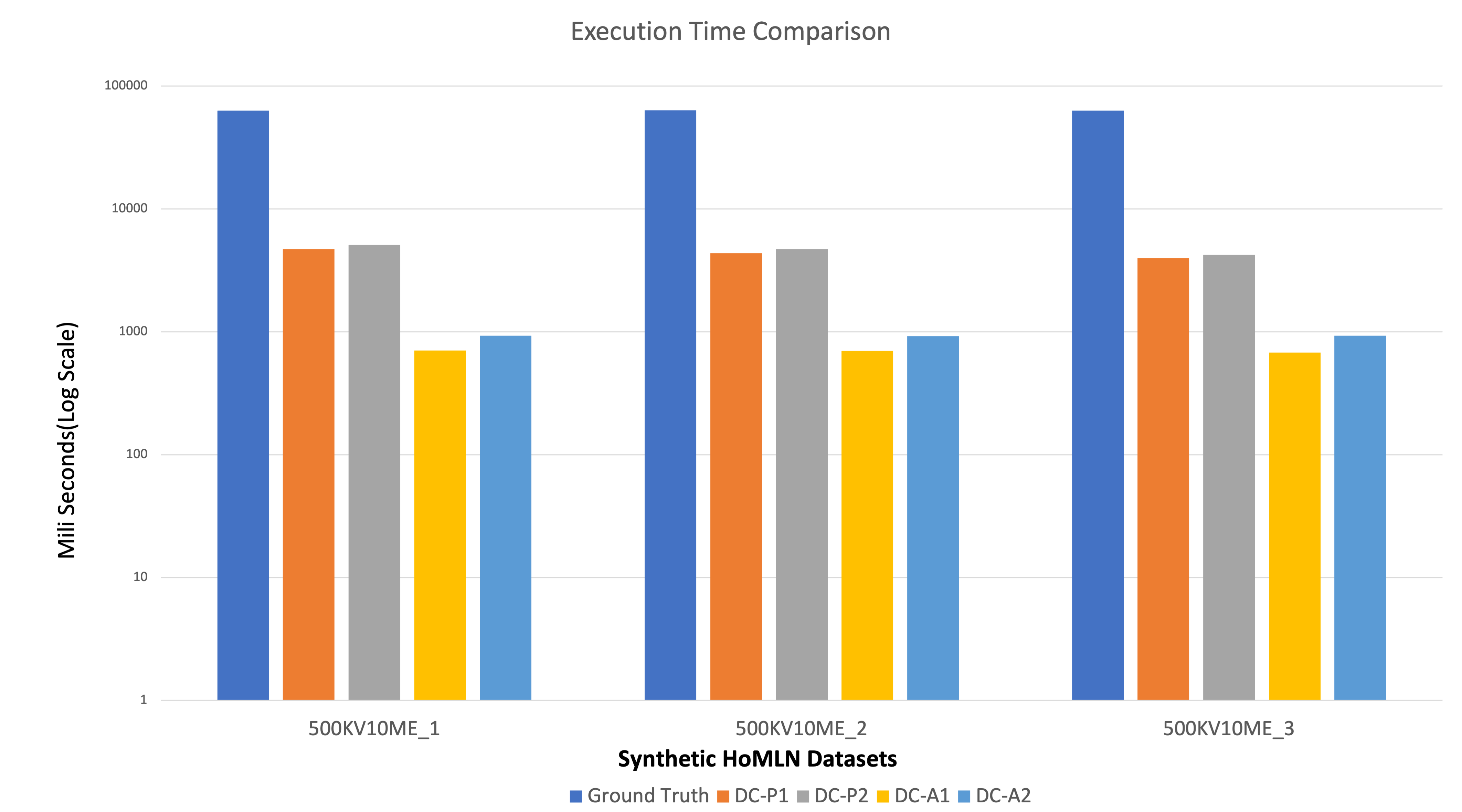, width = 0.5\textwidth}}
\caption{Comparison of Execution Time of the Heuristics against Execution Time of Ground Truth.}
\label{fig:exectime_syn1}
\end{figure}

 When precision is used as the measure of accuracy, DC-P1 and DC-P2 outperform DC-A1, DC-A2, as well as the naive approach. For the synthetic data sets, the precision of DC-P1 and DC-P2 is \textbf{always 100\%} \ref{fig:precision_syn1} and more than 96\% for the real-world data sets \ref{fig:precision_realworld}.
 Figure \ref{fig:exectime_syn1} shows the comparison of the execution time of our proposed solutions against the ground truth time for 3 of the \textbf{largest HoMLN of the synthetic data set 1}. The execution time of our approach is calculated as \textit{maximum $\Psi$ time of the layers + $\Theta$ time}. The ground truth time is computed as \textit{time required to aggregate layers into a single graph using Boolean OR function + time required to find the degree hubs of the aggregated graph}.
 
\textbf{As we can see from Figure \ref{fig:exectime_syn1}, ground truth execution time is more than an order of magnitude as compared to our proposed approaches in all cases (plotted on log scale)}. 
 
 As previously mentioned in Section \ref{sec:mln_decoupling_approach}, it is a challenge to identify and keep the minimum amount of information required in the network decoupling approach. Theoretically, as more information is kept, the accuracy should go up. For this demonstration, we used a HoMLN consisting of 100K nodes from the synthetic dataset 2 where the first layer follows the power-law distribution and the second layer follows the normal distribution. This HoMLN was taken to minimize any similarity among the layers. The additional information kept is the one-hop neighbors of the nodes in each layer. 
 
\begin{figure}[htb]
  \centering
   {\epsfig{file = 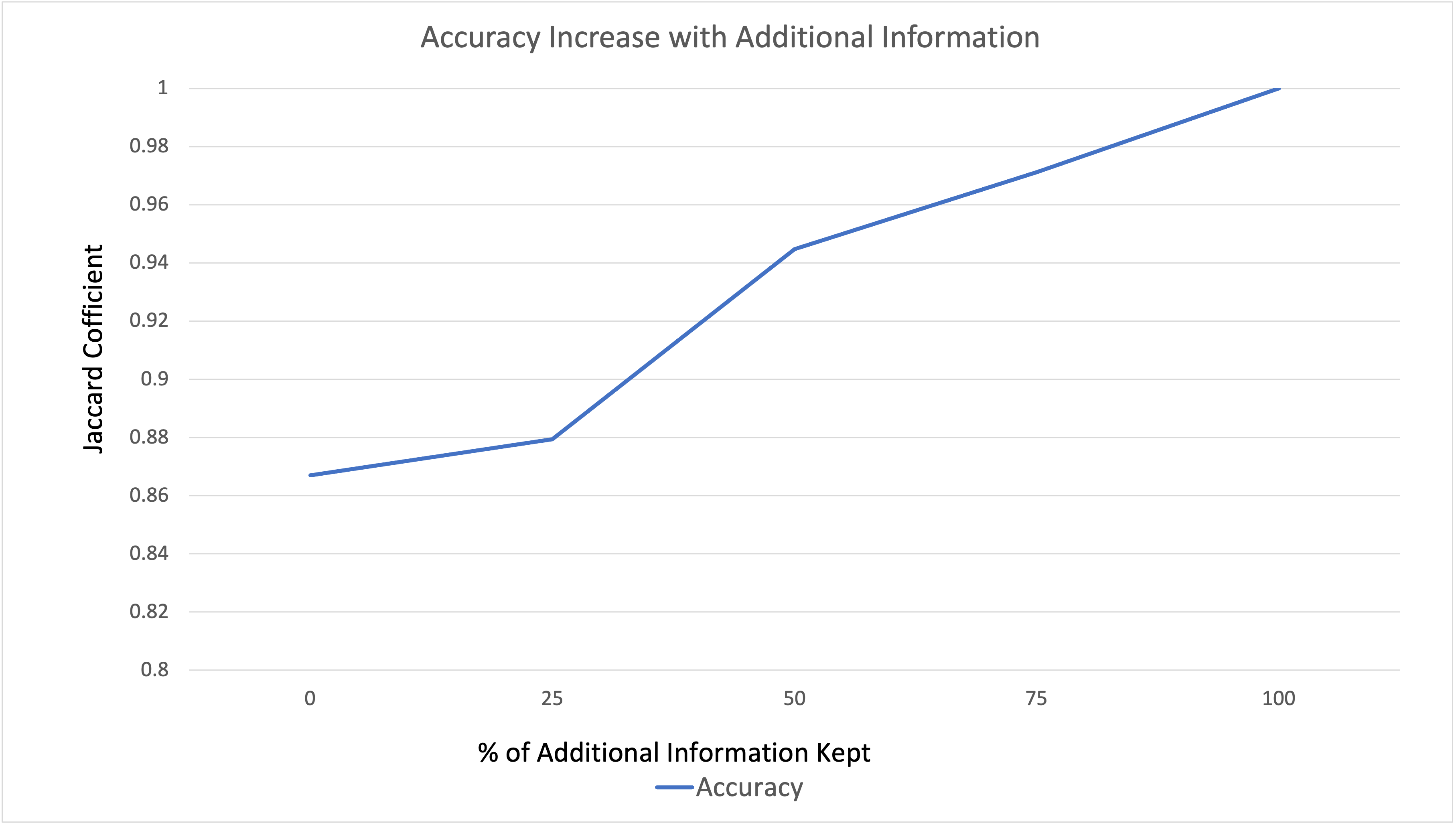, width =0.5\textwidth}}
  \caption{Demonstration of increase in accuracy as more information is kept in each layer for DC-A2.}
  \label{fig:accuracy_additional_info}
 \end{figure}

 In Figure \ref{fig:accuracy_additional_info} we show that as more information is kept in each layer, the accuracy increases. If no information is kept, we get the lowest accuracy. If we keep one-hop neighbors of \textbf{all} the nodes, we get 100\% accuracy.

\section{Conclusions and Future Work}
\label{sec:conclusions}

In this paper, we proposed several heuristics-based algorithms to compute degree hubs in a HoMLN \textit{directly using the decoupling approach}. Some of the heuristics (DC-A1 and DC-A2) achieve high accuracy whereas others (DC-P1 and DC-P2) achieve a precision of 1.
All proposed algorithms show \textbf{more than an order of magnitude improvement in efficiency} as compared to the traditional aggregation approach used for ground truth. Our hypothesis with respect to more information leading to higher accuracy is also established. Future work includes understanding the cascading effects of accuracy and precision when more layers are used. Also, how to identify and retain additional information that can be used to improve the accuracy of multiple layer centrality computation.

%%SC:Though our proposed heuristic DC4 is faster and has higher accuracy compared to the naive approach, our initially proposed heuristics has a precision of 1. We also show that, as more information is kept, the accuracy of our heuristic increase. As of now, our heuristics only work with HoMLN with undirected layers. In the future, we can extend our heuristics to support directed layers. We can also extend the decoupling-based approach to other centrality measures such as betweenness centrality and eigenvector centrality.

%%\vfill
%%\section*{\uppercase{Acknowledgements}}

%%If any, should be placed before the references section
%%without numbering. To do so please use the following command:
%%\textit{$\backslash$section*\{ACKNOWLEDGEMENTS\}}

\bibliographystyle{apalike}
{\small
\bibliography{./bibliography/pavelResearch,./bibliography/santraResearch}}

%%\section*{\uppercase{Appendix}}

%%If any, the appendix should appear directly after the references without numbering, and not on a new page. To do so please use the following command:
%%\textit{$\backslash$section*\{APPENDIX\}}

\end{document}